\documentclass[journal,compsoc,review]{IEEEtran}

\usepackage{float}
\usepackage{amssymb}
\usepackage{amsmath}
\usepackage{mystyle}
\usepackage{graphicx}
\usepackage[]{algorithm2e}
\usepackage{etoolbox,lipsum}
\usepackage{xcolor}
\usepackage{booktabs} 
\usepackage{csquotes}
\usepackage{hyperref}

\DeclareGraphicsExtensions{.pdf,.png,.jpg,.eps}
\DeclareRobustCommand*{\IEEEauthorrefmark}[1]{%
  \raisebox{0pt}[0pt][0pt]{\textsuperscript{\footnotesize\ensuremath{#1}}}}
\graphicspath{ {./imgs/} {./} }

\ifCLASSOPTIONcompsoc
  \usepackage[nocompress]{cite}
\else
  \usepackage{cite}
\fi

\begin{document}

\title{Spatio-temporal Manifold Learning for Human Motions via Long-horizon Modeling}

%
%

\author{He~Wang \IEEEauthorrefmark{1}
        Edmond~S.~L.~Ho \IEEEauthorrefmark{2}
        Hubert~P.~H.~Shum \IEEEauthorrefmark{2} and
        Zhanxing Zhu \IEEEauthorrefmark{3}
\IEEEcompsocitemizethanks{\IEEEcompsocthanksitem 1. H. Wang is with the University of Leeds, Leeds, UK.\protect\\
E-mail: h.e.wang@leeds.ac.uk. ORCID:orcid.org/0000-0002-2281-5679
\IEEEcompsocthanksitem 2. Edmond S. L. Ho and Hubert P. H. Shum are with Northumbria University, Newcastle Upon Tyne, UK.
\IEEEcompsocthanksitem 3. Zhanxing Zhu is with Peking University and Beijing Institute of Big Data Research, Beijing, China}
}

\IEEEtitleabstractindextext{
\begin{abstract}
Data-driven modeling of human motions is ubiquitous in computer graphics and computer vision applications, such as synthesizing realistic motions or recognizing actions. Recent research has shown that such problems can be approached by learning a natural motion manifold using deep learning on a large amount data, to address the shortcomings of traditional data-driven approaches. However, previous deep learning methods can be sub-optimal for two reasons. First, the skeletal information has not been fully utilized for feature extraction. Unlike images, it is difficult to define spatial proximity in skeletal motions in the way that deep networks can be applied for feature extraction. Second, motion is time-series data with strong multi-modal temporal correlations between frames. On the one hand, a frame could be followed by several candidate frames leading to different motions; on the other hand, long-range dependencies exist where a number of frames in the beginning correlate to a number of frames later.  Ineffective temporal modeling would either under-estimate the multi-modality and variance, resulting in featureless mean motion or over-estimate them resulting in jittery motions, which is a major source of visual artifacts. In this paper, we propose a new deep network to tackle these challenges by creating a natural motion manifold that is versatile for many applications. The network has a new spatial component for feature extraction. It is also equipped with a new batch prediction model that predicts a large number of frames at once, such that long-term temporally-based objective functions can be employed to correctly learn the motion multi-modality and variances. With our system, long-duration motions can be predicted/synthesized using an open-loop setup where the motion retains the dynamics accurately. It can also be used for denoising corrupted motions and synthesizing new motions with given control signals. We demonstrate that our system can create superior results comparing to existing work in multiple applications.
\end{abstract}

}

\maketitle


\begin{figure}
    \centering
    \includegraphics[width=0.98\linewidth]{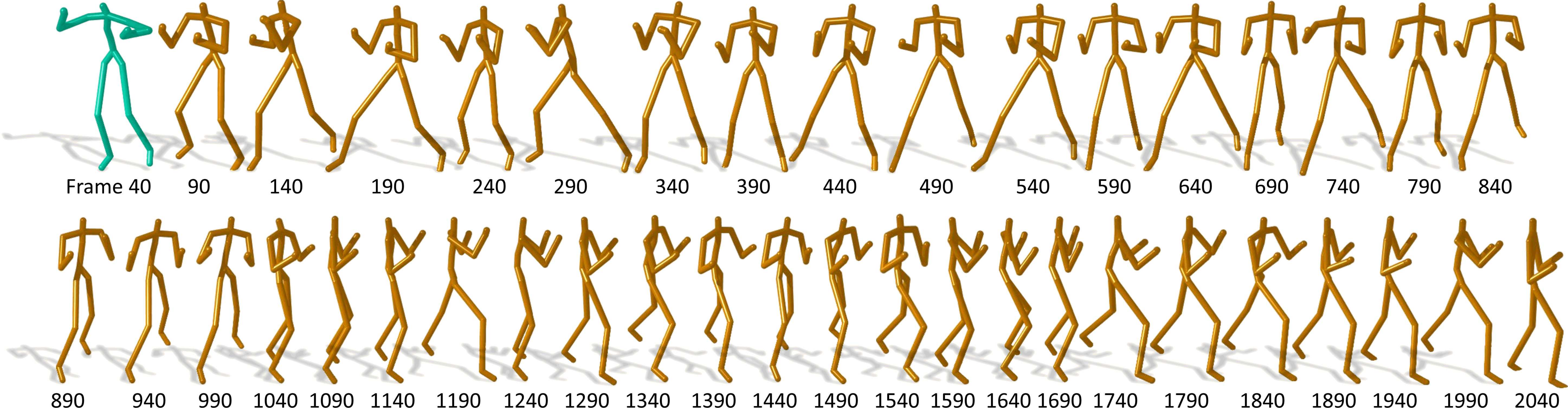}
    \caption{Long-horizion motion generation: given the first {\bf20} frames, STRNN generates the next {\bf20000} frames (yellow) in an open-loop setting. Only the first 2000 frames are shown here.}
    \label{fig:teaser}
\end{figure}


%

\section{Introduction}
\label{sec:introduction}

Modeling natural human motions is a central topic for data-driven character animation. It has been argued that natural human motions constitute a motion manifold \cite{Holden:SIGGRPAH2016}. Although there is a large body of research attempting to find good representations for this manifold such as Finite State Machines \cite{Kovar:2002:MG}, subspace modeling \cite{Taylor:2006} and statistical modeling \cite{min:SIGGRAPH2012}, it is still challenging to find a representative subspace that leads to high-quality modelling.


Recently, some successes have been shown in obtaining a motion manifold using deep learning \cite{Holden:SIGGRPAH2016,Holden:SIGGRPAH2017,Fragkiadaki:2015}, but designing a good network is difficult. The difficulties lie in the spatial and temporal variances of the motion data. Spatially, unlike images, human motions are parameterized on a graph structure (i.e. the skeleton). The local similarity assumption, upon which powerful networks are built for feature extraction, is very different on skeletons comparing to images. The lack of spatial feature extraction leads to the system converging to a \enquote{mean posture} in which the spatial variation of the body movement disappears, as suggested by \cite{Fragkiadaki:2015}. To mitigate such problems, Holden et al. \cite{Holden:SIGGRPAH2016} employed different \textit{disambiguation networks} to recover different types of motions. Temporally, motions show two layers of variances, the first being multi-modality (i.e. the same motion prefix could lead to multiple motions); the second being dynamics variations (e.g. the same set of postures in a motion but with different timing profiles). When generating motions, under-estimating the variances results in a system that generates motions with ``mean posture'' \cite{Fragkiadaki:2015}, while over-estimating them leads to jittery motions. Previous work partially tackles this problem by switching the state of the network \cite{Holden:SIGGRPAH2017} based on the foot contact information during locomotion, but it is difficult to generalize to different types of motions.

We propose a new network called Spatio-temporal Recurrent Neural Network (STRNN) to model the spatial and temporal variances. In particular, we use networks to model segmented parts of the human skeleton to model the spatial variations within a frame effectively. This avoids generating a mean posture and thus eliminates the needs of a separate disambiguous network as in \cite{Holden:SIGGRPAH2016}. Our temporal network performs batch encoding, decoding and prediction, allowing us to use a new loss function to consider long-horizon prediction error and motion naturalness. As a result, our network can learn a high-quality motion manifold that can synthesize long sequences of motions without explicitly introducing extra temporally-related variables as in \cite{Holden:SIGGRPAH2017}. 

To demonstrate the quality and versatility of the learned motion manifold, we first synthesize long, natural and highly dynamic motions using an open-loop setup, in which we do not moderate the error using any run-time systems or disambiguous networks. Then, we show how the manifold can be used to denoise corrupted motion data and synthesize new motions with control signals. We also compare our methods qualitatively and quantitatively with existing methods. 

The proposed network creates high-quality motion manifolds for a variety of applications in both computer graphics, including motion synthesis, motion denoising and motion prediction/extrapolation.

The major contributions of the work can be summarized as follows:
\begin{itemize}
\item We propose a new deep learning framework for generating a high-quality manifold of 3D skeletal human motion.
\item We propose a spatial model for 3D motion by dividing the skeleton structure into parts with spatial proximity and semantics.
\item We propose a new optimization strategy considering long-horizon prediction in the temporal domain to preserve long-term motion naturalness and dynamics.

\end{itemize}

The paper is arranged as follow. We review existing research in Section \ref{sec:related}, particularly focusing on deep learning approaches. Then, we explain how we prepare training data in Section \ref{sec:datapre}. We detail the design of our deep learning model and explain our optimization strategy in Section \ref{sec:method}. We present experimental results of different applications and system evaluations in Section \ref{sec:exp}. Detailed comparisons are shown in \secref{comparison}, followed by the performance analysis in \secref{performance}. Finally, we discuss our system in Section \ref{sec:discussion} and conclude this research in Section \ref{sec:conclusion}.
\section{Related Work}
\label{sec:related}


\subsection{Data-Driven Animation}
Traditional human motion synthesis and classification are done in a feature space which is typically represented as a temporal sequence of 3D skeletal postures \cite{Lee:2002:ICA:566654.566607}. These systems allow some control over the motion, but both the data representation and control schemes are usually manually designed and application specific \cite{Choi:2003:PBL:636886.636889}.

Later, researchers started to look for compact motion representations by statistical modeling \cite{min:SIGGRAPH2012}, dimensionality reduction such as Gaussian Process Latent Variable Model \cite{Grochow:2004:SIK:1015706.1015755} or representative posture landmarks \cite{5291684}. High-level features such as styles are also modeled by, for instance, spectral-domain representation by Fast Fourier Transform \cite{Yumer:2016:SST:2897824.2925955} or local autoregression models \cite{Xia:SIGGRAPH2015}. A similar idea is proposed to represent the formation of a group of characters with a spectral-based manifold for formation interpolation \cite{CGF:CGF1404}. To construct a posture space with a uniform density, selective resampling on a collection of human motion data is proposed \cite{Yang:2015:CDS:2787104.2787132}.

However, due to the non-linearity of human motion, it is difficult to create a global manifold. As a result, local models assuming local linearity are proposed by considering only the data samples that are relevant to the control signals \cite{Liu:2011:RHM:1944745.1944768}. PCA has shown to be effective in reducing a motion into a low-dimensional space for better control and visualization \cite{Safonova:2004:SPR:1015706.1015754,Shin:2006:MSE:1144457.1144464}. It is also applied to a dynamic set of samples selected based on the control signals to construct a low-dimensional space for motion synthesis \cite{Chai:2005:PAL:1073204.1073248}. Weighted PCA with a hierarchical k-d tree structure enables the local models to represent a large database \cite{Tautges:2011:MRU:1966394.1966397}. Considering the reliability of the control signal, a more accurate set of data samples can be found in \cite{shum13realtime}. A mixture of Gaussian process is used to reduce the database size required for motion synthesis \cite{liu16kinect}. These methods (\cite{shum13realtime,liu16kinect}) focus on removing the noise from captured motions, which is a problem known as motion denosing \cite{5406515}. We also demonstrate how our generated manifold can be used for motion denosing with superior results.

Generating local models require high run-time overhead and the accuracy depends heavily on how the local data is selected. In this paper, we are interested in deep learning based approaches, which has the advantage of low run-time overhead as a global model but also model the motion manifold as good as a local models.

\subsection{Deep Learning}
3D skeletal human motion is an effective representation for motion synthesis and classification, in which the movement is presented as temporal sequences of joint positions or joint angles. While deep learning methods using convolutional neural network (CNN) are effective for images \cite{Krizhevsky:2012:ICD:2999134.2999257}, it is unclear how CNN can be performed for skeletal motion as the feature space does not explicitly encode proximity. Existing work shows CNN applied in the temporal domain through an autoencoder that learns the temporal features of a human motion \cite{Holden:SIGGRPAH2016}. The motion manifold learned by the new framework can be used for synthesizing new motions base on high-level parameters. Such an autoencoder, however, does not encode the spatial information of the joints within a frame, leading to the generation of featureless mean postures. A second disambiguation network is required to restore the motion. Another solution is to explicitly encode extracted control parameters such as stepping patterns and stepping phases \cite{Holden:SIGGRPAH2017}, such that the network can output both motion and control parameters for better motion generation. However, such a method is application dependent and cannot be generalized to different types of motions. It is also possible to apply deep learning to learn the control mechanism of a dynamic controller to create dynamic locomotion on different terrains \cite{2017-TOG-deepLoco}.

Recurrent Neural Networks (RNN) has been applied to modeling (\cite{MartinezBR_corr17,AutoCondition}) and recognizing 3D skeletal human motion \cite{Du:RNN:CVPR2015}. In \cite{Du:RNN:CVPR2015}, the skeleton in each frame is divided into different body parts to construct a hierarchical structure of RNN. We follow this idea but designed a network to explore spatial features of the body parts. Modeling temporal information using deep neural network has also been applied in handling RGB videos. Recurrent Convolutional Neural Networks (RCNN) are proposed to handle video data, in which each frame is fed into a CNN followed by one or more layers long short-term memory (LSTM) that are recurrently connected \cite{Donahue:LSTM:PAMI2016}. A similar RCNN structure is applied for human identification using 4D depth video, in which a spatial distribution of 3D point cloud obtained by a single depth is used at the input of the CNN \cite{Haque:CVPR2016}. A recent supervised learning approach \cite{Lee:SIGGRPAHAsia2018:CharCtrl} is proposed for synthesizing character animation with interactive control. In particular, the motion synthesis framework models the spatio-temporal motion structure as well as constraints for interactive control by RNN-based networks. While we share similar interests in using RNN for synthesizing high-quality human motion, our research is orthogonal to theirs. In \cite{Lee:SIGGRPAHAsia2018:CharCtrl}, they trained the model for the tasks using task specific objectives. In contrast, our method aims to learn a versatile and natural motion manifold without any task-specific supervisory information. We propose a RNN model equipped with a spatial network, allowing it to directly utilize skeletal information of human motions to obtain a high-quality manifold.

Spatio-temporal graphs that explicitly represent the relationships of human and environment objects have shown to be effective in activity recognition \cite{Li2008}. Structural RNN can be used to recognize activities based on graphs representing the trajectories of skeletal motions and object movements \cite{Koppula:2013}. Another graph implementation is to consider the spatial and temporal relationship as intra-frame and inter-frame edges respectively \cite{Li2008}. However, these graphs focus more on the interaction between multiple instances, instead of focusing on the movement of a single human. Also, they are not suitable for motion synthesis due to the abstract representation.
\section{Data Preparation}
\label{sec:datapre}
We follow the practice in \cite{Holden:SIGGRPAH2016} to build our motion dataset. We use several datasets including CMU \cite{CMU}, HDM05 \cite{HDM05}
, MHAD \cite{Ofli:WACV2013}, Action3D \cite{Li:CVPR2010} and Edinburgh \cite{Holden:SIGGRPAH2016}. Since the captured data comes from different actors with different skeletons, the data is first scaled then mapped onto one standard skeleton. Then inverse kinematics is used to bring the joints of the standard skeleton to the joint positions of the source skeleton. Frame rates from different datasets vary from 30 to 120Hz. We resampled them to 30Hz. To unify posture presentations, we use joint positions defined with respect to the body's local coordinate system where we project the root onto the ground as the coordinate system origin. Some other methods used joint angles instead joint positions. We also tried joint angles but found that it is easier for neutral nets to learn a stable manifold of joint positions. Also, it is also easier for motion control, explained in the next section. Our character contains a total of 73 Degrees of freedom (Dofs). The first 66 Dofs are the joints including 6 Dofs (the global positions and orientation) of the root joint and 3 Dofs for the 3D positions of the remaining 20 joints (left/right toes, foot, knees, hips, fingers, wrists, forearms, arms and spine, spine1, neck, head). Then we also have a global velocity which is a 2D vector in the x-z plane (assuming that the y axis points upwards) and an angular velocity which is a 1D vector around the y axis. Finally, we use 4 binary variables to record binary foot contact information for left heel, left toes, right heel and right toes respectively.  After processing, we obtained approximately 60k motion clips that are further divided in ratio 80:10:10 for training, validation and testing.
\section{Spatio-temporal Recurrent Neural Network (STRNN)}
\label{sec:method}

\subsection{Motion Manifold Modeling}
We start by parameterizing the motion manifold as a time series: $P(X_{t+n}, ..., X_{t+1} | X_t, ..., X_{t-m})$ where $X_t$ is the motion frame at time $t$ and $P$ is the conditional probabilistic distribution of $n$ frames from $t+1$ given $m+1$ frames before $t+1$. What the model captures is the dependencies between the past $m+1$ frames and the future $n$ frames. Many existing data-driven models fall under this umbrella. Most of them consider the situation when $n=1$ and $m=0$, such as the motion graphs \cite{Kovar:2002:MG} and autoregression \cite{Xia:SIGGRAPH2015}. Some consider $m>0$, such as the dynamic Bayesian networks \cite{Lau:SIGGRAPH2009} and high-order Markovian models \cite{Wang:PAMI2008}. Our model generalizes them by setting both $n$ and $m$ much bigger than 1, resulting in a \textit{batch prediction} framework that forces the model to look forward into a further future and capture the mid-term and long-term frame dependencies. This allows us to improve the model representation of the motion dynamics. Here, we propose the Spatio-temporal Recurrent Neural Network (STRNN) that learns these dependencies.

\begin{figure*}
    \centering
    \includegraphics[width=.8\linewidth]{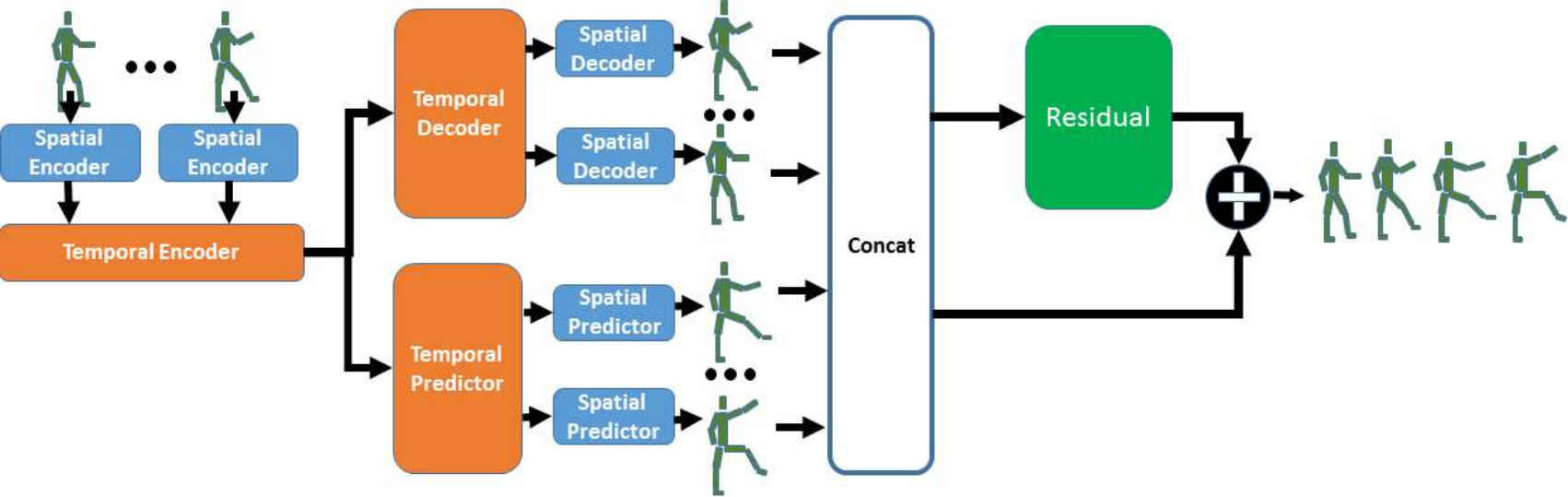}
    \caption{The network architecture of the Spatio-temporal Recurrent Neural Networks. Detailed network structures of  the Temporal Encoder/Decoder/Predictor and the Spatial Encoder/Decoder/Predictor can be found in Fig. \ref{fig:tempNet} and Fig. \ref{fig:spatialNet} respectively.
    }
    \label{fig:STRNN}
\end{figure*}

\subsubsection{Spatio-temporal Recurrent Neural Network (STRNN)}
STRNN consists of three sub networks: spatial, temporal and residual, shown in \figref{STRNN}. The temporal network aims at learning the temporal dependencies between long motion sequences. It is structured as a Recurrent Neural Network composed of long short-term memory (LSTM) cells \cite{Hochreiter:1997}. 

\begin{figure}
    \centering
    \includegraphics[width=.8\linewidth]{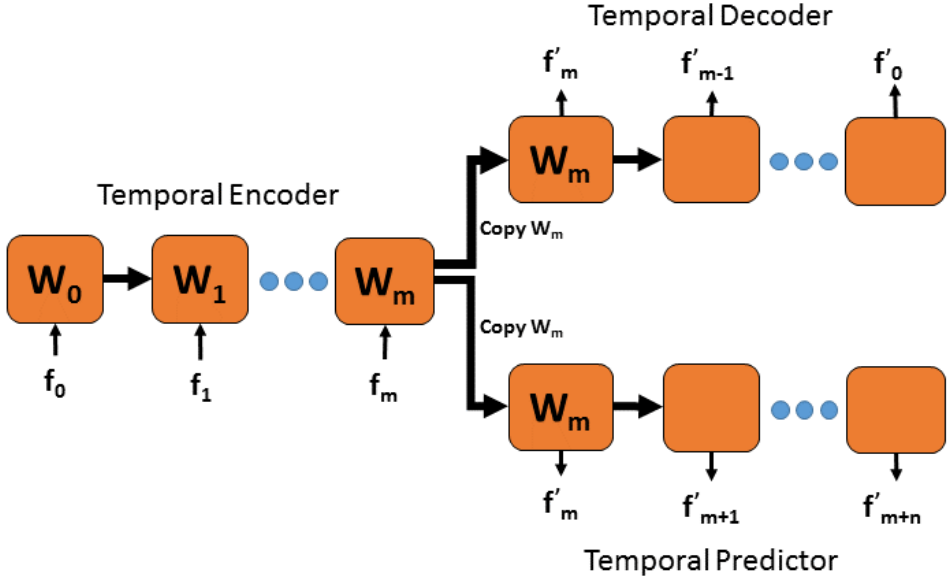}
    \caption{Two-way Bidirectional Temporal Network (TBTN). Rectangles are LSTM cells. $W$s are the corresponding weights.}
    \label{fig:tempNet}
\end{figure}

LSTM networks have been used for modeling time series data in many fields \cite{Donahue:LSTM:PAMI2016}. Two common approaches are (1) encoding/decoding \cite{Zhang:CoRR2017} where information is forced to be recovered by every time step, and (2) sequence-to-sequence \cite{SutskeverVL:Corr2014} where the network takes all input first, then decodes them into a different time series. The former focuses on learning the inter-frame dependencies while the latter targets at the mappings between sequences. Our temporal network is a combination of both, so that the network is forced to reconstruct the motion manifold while taking the future motion into consideration, as shown in \figref{tempNet}. It also enables us to impose constraints in a longer time span to stabilize the network.

The temporal network is named \textit{Two-way Bidirectional Temporal Network} (TBTN), consisting of three parts: the temporal encoder (TEncoder), the temporal decoder (TDecoder) and the temporal predictor (TPredictor) (\figref{tempNet}). The training is done by iterations of forward and backward passes. The forward pass goes through an encoding phase and then a decoding/predicting phase. It starts by taking $m+1$ frames into TEncoder. Note that unlike some RNNs \cite{Fragkiadaki:2015}, the decoding and predicting only start \textit{after} the TEncoder takes all the input, making it a sequence-to-sequence model. After the encoding phase, the internal state of TEncoder is copied to TDecoder and TPredictor as a \textit{good/reasonable initialization}. Then, the forward pass continues on TDecoder and TPredictor simultaneously. The decoding in TBTN unrolls in both directions in time. The task of TDecoder is to decode the frames \textit{backwards} in time and the task of the TPredictor is to predict the frames \textit{forwards} into the future. The backward decoding improves the convergence speed as it first decodes the last few frames that the encoder just sees. Therefore, it forces the model to learn the short-term correlations between frames first, and then progresses onto their longer-term correlations. Similar ideas have been employed in video prediction problems \cite{Srivastava:CoRR2015}. During the decoding phase, the decoded frame $f'_{i}$ serves as the input for decoding $f'_{i-1}$. Similarly, the predicted frame $f'_{j}$ serves as the input for predicting $f'_{j+1}$. The forward pass of TEncoder is defined as:
\begin{equation}
\label{eq:lstm}
\Phi_e(X) = LSTM_e(W_eX + b_e)
\end{equation}
where $X = \{f_0, f_1, ..., f_{k-1}\}$ is a motion sequence with $k$ frames, $(W_e, b_e)$ are are the weights and biases of the LSTM cells in TEncoder. Similarly, we use $\Phi_d$, $\Phi_p$, $(W_d, b_d)$ and $(W_p, b_p)$ to represent TDecoder and TPredictor respectively. 

Although TBTN can learn the temporal dependencies, the representation of the frame itself also requires a modeling strategy. The simplest solution is to feed in raw data such as joint positions and joint angles. However, human motions are highly coordinated \cite{Grochow:SIGGRAPH2004:SIK} while the raw representation ignores the correlations. To learn the correlations directly from raw data, it requires a huge amount of data with large variety. In addition, even if the data is available, the variance can easily be overestimated or underestimated by the RNN networks \cite{Fragkiadaki:2015}.

\begin{figure}
    \centering
    \includegraphics[width=.5\linewidth]{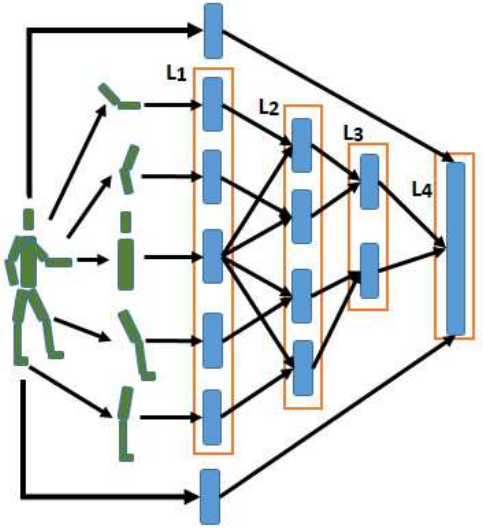}
    \caption{Spatial encoder. A hierarchical neural network on graph structure.}
    \label{fig:spatialNet}
\end{figure}

To capture the correlations, we observe that the Dofs often move as relatively independent groups based on the semantics of the motion. For example, waving a hand mainly involves the Dofs on one arm. While this kind of group independence may not be universal in all motions, grouping Dofs provides better modelling of the correlations between groups and the overall body in general \cite{Du:RNN:CVPR2015}. Based on this observation, we propose to perform hierarchical spatial modeling on each set of body part(s), as shown in \figref{spatialNet}. At L1, each group-wise component generates a \textit{summary} for the whole group. For instance, the L1 component for the left arm encodes the variances of all the left arm Dofs in different motions. When the summaries at L1 are combined at L2, the cross-group correlations are modeled. The same principle applies up to L4. Overall, to encode a frame, we group DoFs based on body parts, and then merge them hierarchically until we have a latent encoding. The decoding/prediction is done reversely. By grouping Dofs, group-wise posture variances and cross-group correlations are explicitly modeled.

The Spatial Encoder (SEncoder) decomposes a human body into seven parts: $X$ = \{$X_{root}$, $X_{torso}$, $X_{lLeg}$, $X_{rLeg}$, $X_{lArm}$, $X_{rArm}$, $X_{fp}$\}. $X_{root}$ includes root positions, velocity and angular velocity around the Y axis. $X_{fp}$ is foot contact information. $X_{torso}$ includes spine, neck and head. $X_{lLeg}$ and $X_{rLeg}$ include the hip, knee and foot of the left/right side. $X_{lArm}$ and $X_{rArm}$ include the shoulder, elbow and hand of the left/right side. We define $X_{body}$ = \{$X_{torso}$, $X_{lLeg}$, $X_{rLeg}$, $X_{lArm}$, $X_{rArm}$\}. Each blue box in the four groups ($L_1$-$L_4$) is a fully-connected layer. The layers in the same group have the same number of hidden units. From $L_1$ to $L_4$, the number hidden units is 64, 128, 256 and 512 respectively. In $L_1$, individual body groups are mapped to a latent space. $L_2$ merges the $lArm$ with $torso$, $rArm$ with $torso$, $lLeg$ with $torso$ and $rLeg$ with $torso$ respectively. $L_3$ merges the body parts into the upper body and the lower body. Finally, $L_4$ merges both the upper and lower bodies into the whole body. Besides, the top component is the global velocity on the x-z plane and the rotation around the y-axis. The bottom component is a 4D binary vector on foot contact. Both components are directly merged into $L_4$, too. We then append a dropout layer\cite{srivastava:JMLR2014} and a batch normalization\cite{IoffeS:CoRR2015} layer to $L_1 - L_4$. The network structure explicitly models the spatial correlations between different Dof groups. The Spatial Decoder (SDecoder) and Spatial Predictor (SPredictor) have the same network structure except that the information is propagated in the reversed direction. We denote the spatial encoder as below:
\begin{eqnarray}
\label{eq:conv}
\Sigma_e(X) = \sigma_{L_4}(\{\sigma_{L_3}(\sigma_{L_2}(\sigma_{L_1}(X_{body})), X{root}, X_{fp}\}) \\
\sigma_{L_x} = BN(drop(tanh(W_s merge(X_{in}) + b_s)))
\end{eqnarray}
where $\sigma_{L_x}$, $x = 1, 2, 3, 4$ are the activation on $L_1$ to $L_4$ layers. $W_s$ and $b_s$ are the weights and the biases of the fully-connected layer. SDecoder and SPredictor decode the latent representation into the original data. They are both denoted by $\Sigma_e^{-1}$. Note our spatial encoder is frame-based, different from \cite{Du:RNN:CVPR2015}. While their aim is to learn the temporal patterns of individual joints, we aim to learn spatial correlations.

We find that the spatial and temporal networks together alone can model motion multi-modality (refer to the accompany video). However, we observe periodic jumps, which is a problem in iterative motion prediction that is also observed in other systems \cite{MartinezBR_corr17}. We formulate the removal of this high frequency noise as a learning problem, i.e. to learn a signal to cancel the noise. To this end, we use a residual component. Our residual component consists of four fully-connected layers with ELU activation functions \cite{ClevertUH15} and 512 hidden neurons, except for the last layer where it maps the data back to the original dimension. Both the input and output of the residual component are motion data (decode\_len + predict\_len, d) where d is the dimension of a single posture, decode\_len and predict\_len are the number frames that are decoded and predicted. The input is a concatenation of the output of the spatial and temporal networks and the output is the filtered results. SDecoder and SPredictor share weights. This is inspired by \cite{Holden:SIGGRPAH2016} that a few full-connected layers can learn a good latent representation. However, note that our solution is very different in the sense that STRNN does not rely on other networks to disambiguate motion modes \cite{Holden:SIGGRPAH2016} or to learn a velocity profile \cite{MartinezBR_corr17}. Instead, the residual component learns the distributions of noises and acts as a filter.

\subsection{Loss Function and Model Training}
Our loss function includes a reconstruction error term. Besides, it contains new terms that are inspired by the past animation research in non deep-learning areas. The first one is the long-horizon costs that governs the smoothness of the generated motions. We found that this seemingly simple cost is the key to harness the motion variances. The second is a group of motion control costs such as control signals and bone-length constraints. Here, we give details of each of the cost terms. We define the full loss function:

\begin{equation}
\label{eq:cost}
Cost(X, \Omega) = \\
w_r C_r + w_s C_s
\end{equation}
where $C_r$ and $C_s$ are reconstruction and smoothness, with weights $w_r$ and $w_s$.

\subsubsection{Reconstruction Error}
We use Mean Squared Error (MSE) for $C_r$ to force STRNN to reconstruct motions. $C_r$ = $C_d$ + $C_p$:
\begin{eqnarray}
C_d = \frac{1}{m}\sum\lVert(\Sigma_e^{-1}(\Phi_d(\Phi_e(\Sigma_e(X_e)))) - X_e)\rVert^2 \\
C_p = \frac{1}{n}\sum\lVert(\Sigma_e^{-1}(\Phi_p(\Phi_e(\Sigma_e(X_e)))) - X_p)\rVert^2
\end{eqnarray}
where $C_d$ and $C_p$ are the reconstruction loss of the decoded and predicted motions. $m$ and $n$ are the decoding and prediction lengths, $X$ = \{$X_e$, $X_p$\}, $X_e$ and $X_p$ are the ground-truth decoding/predicted motions, and
$\Sigma_e^{-1}$ is simply the inverse function of $\Sigma_e$.  Minimizing $C_r$ results in a tight approximation of the motion manifold. We will denote this cost as \textit{MSE} in Section \ref{sec:exp}. 

\subsubsection{Long-horizon (LH) Cost}
$C_s$ in \eqref{cost} is smoothness cost:
\begin{eqnarray}
\label{eq:smoothness}
C_s = &\frac{1}{m+n}\sum \lVert \hat{X}^{t+1}_{body} - 2\hat{X}^{t}_{body} + \hat{X}^{t-1}_{body}\rVert^2 \nonumber \\
& + \sum \lVert \hat{X}^{t}_{root} - \hat{X}^{t-1}_{root} \rVert^2
\end{eqnarray}
where $m$ and $n$ are the decoding and prediction lengths, $\hat{X}$ is the concatenation (along the time axis) of the decoded and predicted motions, $C_s$ governs the smoothness of the motions and has been widely used in many optimization-based character animation approaches. Note this constraint essentially penalizes big accelerations only and does not overly dampen the motion dynamics. We will denote it as \textit{long horizon constraint (LH)} in Section \ref{sec:exp}.

\subsubsection{Run-time Costs}
\label{sec:runtimeCosts}
Since the root velocity, root angular velocity around Y-axis and foot planting information are also included in the training data, control can be imposed in run time by penalizing the deviations of the decoded and predicted motion from the desired motion signals. These penalties can be added to \eqref{cost} with a weight equal to ($1-w_r-w_s$): 
\begin{equation}
    C_{con} = H_{ctr} + H_{bone} + H_{fp}
\end{equation}
where:
\begin{eqnarray}
    H_{ctr} = \sum \lVert \hat{X}_{root} - \Gamma \rVert^2 \\
    H_{fp} = \sum f_{cta} \Vert\hat{X}'_{foot}\rVert^2 \\
    H_{bone} = \sum \lVert \lVert \hat{X}_{body}^{i} - \hat{X}_{body}^{j} \rVert - l^{ij} \rVert^2
\end{eqnarray}
$H_{ctr}$, $H_{bone}$ and $H_{fp}$ are costs on control signals, bone length and foot contact respectively. $H_{ctr}$ ensures the control is satisfied. $\Gamma$ is the given control signal specifying root velocity and root angular velocity around the y-axis, allowing us to safely recover the global translation and rotation. $H_{fp}$ fixs the foot sliding by penalizing the foot velocity $\hat{X}'_{root}$ where $f_{cta}$ is the stepping patterns. $f_{cta}$ = 1 if there is foot contact and 0 otherwise. It can be extracted from the motion or manually specified. We use a simple heuristic to automatically detect foot contact states by thresholding the heights and the world speeds of the toes and heels. $H_{bone}$ enforces the bone-length constraints with $l^{ij}$ being the bone length between joint $i$ and $j$.  

\subsubsection{Training}
\label{sec:training}
As long as \eqref{cost} is differentiable, we can use stochastic gradient decent for optimization. In our experiments, AdaDelta \cite{Zeiler:CoRR2012} is used with random weight initialization. We set learning rate = 1, $\rho$ = 0.95, $\epsilon$ = 1e-08 and decay = 0.0. 

To accelerate training and prevent overfitting, first, we regularize $W$s and $b$s on their L2 norms in \eqref{cost}, which forces the model to use as few activations and biases as possible, both weight coefficients are set to 0.01. Second, we use Dropout \cite{srivastava:JMLR2014} with 10\% rate and batch normalization \cite{IoffeS:CoRR2015} after each fully-connect layers in the spatial network.

Next, we use corrupted inputs to train STRNN. Randomly corrupted inputs have been found effective in training \cite{Fragkiadaki:2015}. Here, we design a special mechanism to corrupt the inputs for different iterations. We use a Gaussian noise with zero mean and 0.1 standard deviation to corrupt the input in the first iteration. We then gradually decrease the deviation by 0.001 in each of the following iterations until the value becomes 0.0. We find this training mechanism beneficial because it first tries to bring the optimizer into a large area centered around the ground truth, and then gradually brings it to the ground truth. On the way, it keeps the memory of the corrupted data, essentially mapping a small area around the ground truth to the ground truth. This helps to reduce error accumulation over time. Unlike \cite{Fragkiadaki:2015} where the noise level gradually increases over time, decreasing the noise level avoids the shortcoming of unable to reliably choose the right model for testing \cite{MartinezBR_corr17}. This is because in the later stage of training, there is no noise in the ground truth, and thus the reported learning and validation errors can be reliably used for selecting the best model.

Finally, we use a hybrid training strategy. We divide the model into the spatio-temporal network (full model without the residual network) and residual network. We first train our spatio-temporal and residual network separately, using LH + MSE and MSE respectively, then compose them together and fine-tune the residual network using only MSE. The spatio-temporal network trained by LH + MSE can generate good motions but with high-frequency noises. Then, the residual network aims to learn to cancel high-frequency noises and the rest aims to learn the motion dynamics. Separate training gives them both good initialization. When fine-tuning, we fix the spatial and temporal networks and only tune the residual component. The reason is that our pre-trained spatial and temporal networks capture the motion dynamics well, but with high frequency and periodic noises (see \secref{exp}). Removing the noise can be seen as learning residuals to cancel them. It is faster for the residual network to learn these residuals if it is pre-trained on the ground-truth.

\section{Experimental Results}
\label{sec:exp}
In this section, we will first evaluate STRNN from several aspects. Next, we will demonstrate three applications, including long sequence synthesis, motion denoising and controlled motion synthesis.

\subsection{Evaluation}

STRNN is a family of neural networks with variations in the structure. One variable is the duration of the temporal network, decided by TEncoder, TDecoder and TPredictor, denoted by \textit{encode\_len}, \textit{decode\_len} and \textit{predict\_len} respectively. We vary the three lengths while enforcing encode\_len = decode\_len. We first evaluate various settings on data consisting of motion segments of \textit{40} frames, denoted by $l=40$ where $l$ is the length of the segments. Then, we empirically find a good setting and test it on data where $l=10$ and $l=20$.
 

To show the added benefits of different components (spatial, temporal and residual), we first show STRNN with only the temporal network, denoted by \textit{Temporal}. Then, we combine the spatial network and the temporal network denoted by \textit{SpatioTemp}. Finally, we show the full model (spatial+temporal+residual) denoted by \textit{Composite}. All evaluations are accompanied with quantitative (errors) and qualitative results (videos) on the task of long-horizon prediction.

We also show the effects of the terms in loss function (\eqref{cost}) and the hybrid training strategy. We use MSE, LH and HY to denote three different settings. MSE means $C_r$ only, LH means $C_r$ and $C_s$. HY is the LH loss with the hybrid training strategy, explained in Section \ref{sec:training}.

In the rest of the section, we use the following model notations. Composite\_20\_20\_HY means we use the full model and hybrid training strategy with 20-frame encoding (i.e. 20-frame decoding) and 20-frame prediction, while Temporal\_39\_1\_MSE means we only use temporal network and MSE loss with 39-frame encoding (i.e. 39-frame decoding) and 1-frame prediction. In addition, all experimental results can be found in the video, in which all the demos (except one for comparison purposes) are not post-processed to show the true performance of our model. We also give one example (\secref{postProcessing}) to show that our model can benefit from incorporating post-processing to remove issues such as foot sliding.

\begin{figure}
\centering
\includegraphics[width=0.98\linewidth]{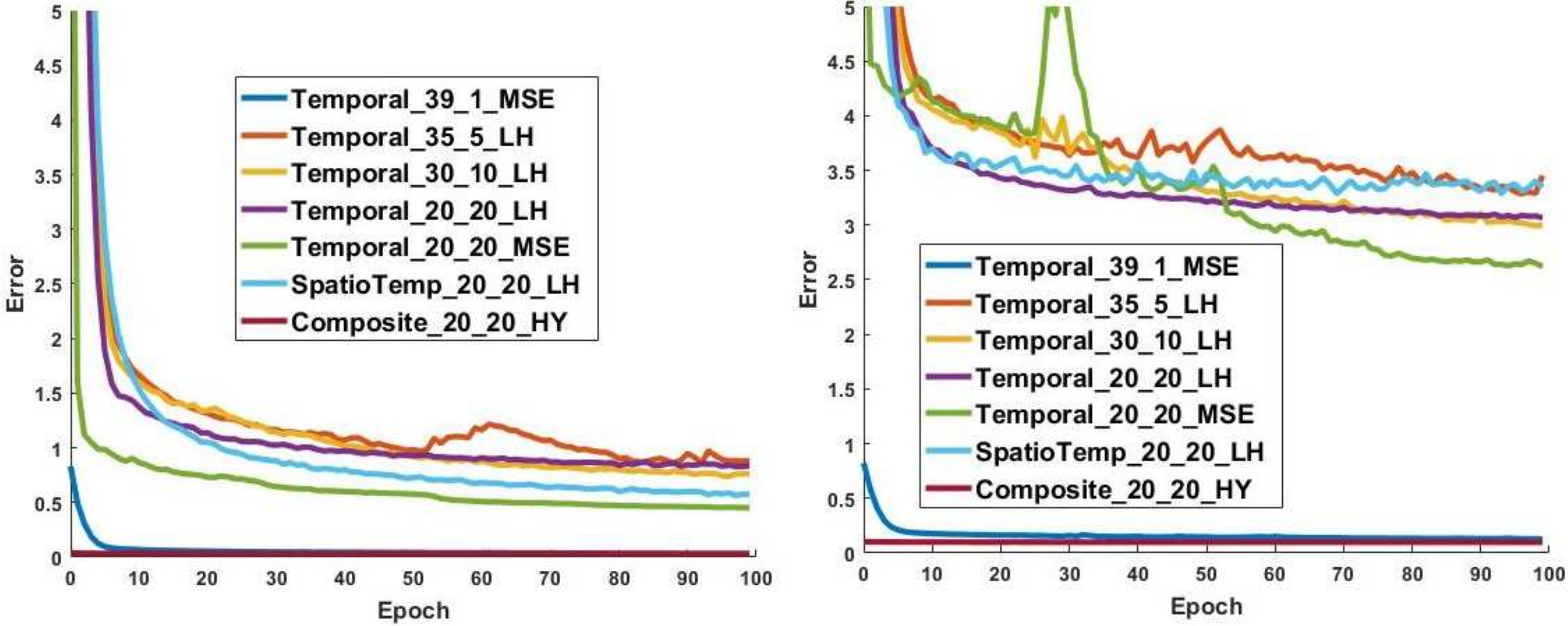}
\caption{From left to right: training error and validation error. SpatioTemp\_x\_y is the model where x is the encode\_len and y is the predict\_len. Temporal\_x\_y is the model \textit{without} the spatial encoder/decoder/predictor and the residual network. Composite\_x\_y is the full model. Also, MSE means only $C_r$ is used. LH means both $C_r$ and $C_s$ are used. HY means the hybrid training results.}
\label{fig:modelChoice}
\end{figure}

\subsubsection{Encoding vs Predicting}
\label{sec:evp}
\figref{modelChoice} shows the decoding and prediction errors for both training and testing phases. We choose encode\_len = 39, 35, 30, 20 for comparison. We emphasize that \figref{modelChoice} only shows that how well different models can fit data on 40-frame motion segments. It is not directly related to the visual quality of generated motions, especially of those longer than 40 frames, explained in \secref{PredictStability}. In term of the data-fitting ability, we find that larger predict\_len generally gives better overall results given $l$ is fixed. The only exception is Temporal\_39\_1\_MSE but visually it generates worse results. This indicates that there should be a balance between the encode\_len and predict\_len. After experiments with different settings, we found that a balanced encoding\_len/predict\_len pair tends to give better results. So, we empirically set encode\_len = predict\_len = 20 for our experiments unless specified otherwise.

\subsubsection{Motion Prediction Quality}
\label{sec:PredictStability}
Next, we do both qualitative and quantitative comparisons on prediction tasks with different network settings. The qualitative visual comparisons can be found in the video
. 
In the quantitative comparison, we found that although \figref{modelChoice} shows how well different models can fit data, the results are not directly correlated to the visual quality. Since designing a metric that directly reflects visual quality is difficult, we use a metric that evaluates how close the generated motions are from the motion manifold, which has been previously used to measure motion style similarity\cite{shum13realtime}. Also, we assume that the existing data are representative samples and are dense enough to represent the 'ground-truth' motion manifold. In practice, we found that this metric is largely consistent with the visual assessment:
\begin{eqnarray}
\label{eq:1nnError}
    D_{1nn} = min(dist(D_g, D_m)) \\
    dist(D_g, D_m) = \frac{1}{n}\sum_{i=1}^{n}{(||\hat{f}_i - f_i||_2^2)} 
\end{eqnarray}
where $D_g$ and $D_m$ are generated and ground truth motions respectively, each with $n$ frames. $\hat{f}_i$ and $f_i$ are in $i$th frame in $D_g$ and $D_m$. $dist(D_g, D_m)$ is essentially the per-frame $l_2$ norm of two motions. $D_{1nn}$ essentially is the smallest distance between $D_g$ and the ground truth data, which is a measure of how close the generated motion is from the manifold. In practice, $D_g$ and $D_m$ could have different duration. An exhaustive comparison is impractical due to the exponential complexity. As a solution, we use a 10-frame overlapping sliding window to chop both $D_g$ and $D_m$ into segments with the same length (40 frames). Then, we compute $D_{1nn}$ by exhaustively comparing all segments in $D_g$ to all in $D_m$. To test the generalizability, we use disjoint training and testing datasets. We use 20-frame motion prefixes randomly selected from the testing dataset to generate 200 frames. Then, we compute the metric $D_{1nn}$.  

\begin{figure}
\centering
\includegraphics[width=.8\linewidth]{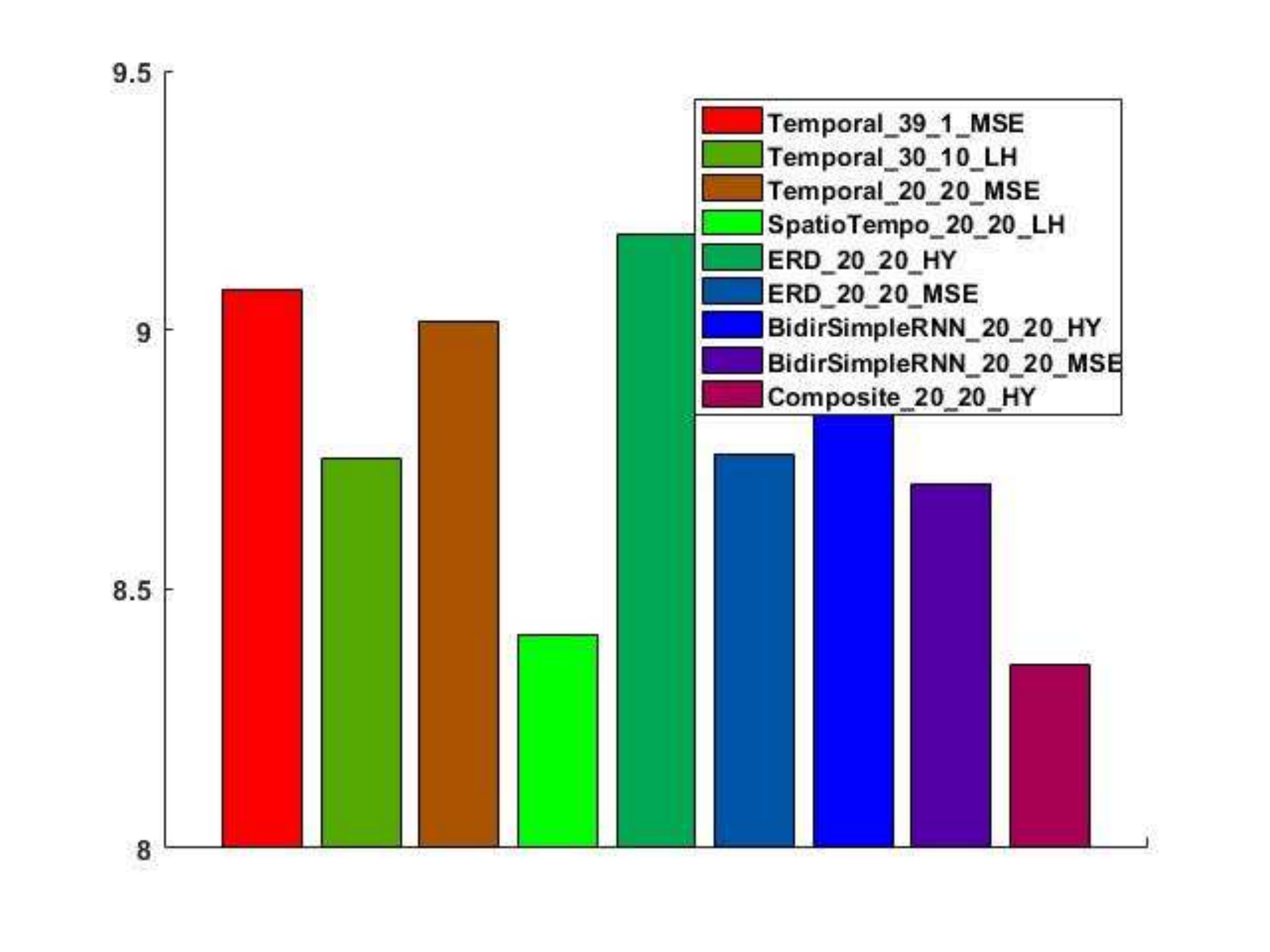}
\caption{The $D_{1nn}$ error of different models on a 200 frame motion prediction task.}
\label{fig:modelComparison}
\end{figure}

The results are shown in \figref{modelComparison}. Our full model (Composite\_20\_20\_HY) outperforms a baseline model (Temporal\_39\_1\_MSE) where no batch prediction is used and no LH loss nor HY training was used (ablation test 1). We also found that LH loss performs better than MSE (ablation test 2). The LH loss only regulates the acceleration, not the position nor velocity, so not overly smoothes the motions. MSE on the other hand has been known to lead to a mean posture as it is an easy local minimum for the optimization. Our speculation is that the combination of both actually allows the optimizer to find a solution that keeps variances so that it does not get stuck in the ‘mean posture’ local minimum. In addition, by testing the network with different components, we also found that batch prediction produces better results in different settings (Temporal\_39\_1\_MSE vs Temporal\_30\_10\_LH vs Temporal\_20\_20\_MSE in \figref{modelComparison}). For the sake of simplicity, a visual comparison on more exhaustive experiments can be found in the video (ablation test 3).

Next, adding the spatial network retains the motion variances better. Sometimes batch prediction can still lead to underestimating the motion variances, observed in Temporal\_20\_20\_LH (ablation test 4). In contrast, SpatioTemp\_20\_20\_LH retains the variances and prevents the mean pose problem, which is a result of group-wise variances being explicitly modeled. 

Further, although SpatioTemp\_20\_20\_LH gives good results (\figref{modelComparison}), it generates periodic jumps that is due to the under-constrained variances near the boundary of predicted motion segments (ablation test 5). Similar observations were also made in \cite{MartinezBR_corr17}. While in \cite{MartinezBR_corr17} the model learns the velocity instead of the posture to eliminate the problem, the side-effect is that it can still converge to the mean pose in a long run. Our residual network combined with our hybrid training strategy not only improves the quality but also retains the variances. Please see the comparison between Composite\_20\_20\_HY and SpatioTemp\_20\_20\_LH (ablation test 5).

\subsubsection{Different Motion Lengths}
The added value of different components have been evaluated above. But they are all done with 40-frame motion segments. We therefore also vary the whole motion segment duration to see if it generalizes well. Given the limit of our graphics card, the longest motion segment we experiment is 40 frames. We therefore experimented with 20-frame and 10-frame data under the setting of Composite\_10\_10\_HY and Composite\_5\_5\_HY (ablation test 6 \& 7 on two different motion prefixes). We found that our network generalizes well. In the video, all of them generate motions with similar visual qualities. This also suggests that strategy of a balanced encoding\_len and predict\_len generalizes well too.

\subsection{Motion Prediction and Extrapolation}
Prediction power is one effective way to test motion manifold models. It is often presented as an initial value problem in which if a prefix of data is given, the model should be able to extrapolate/predict into the future while keeping the predictions on the manifold.

After training our network on the CMU dataset, we take 20 frames as input and let STRNN predict 20 frames into the future. STRNN can reliably predict the next 20 frames. Then, we push STRNN further to extrapolate for even longer sequences to test the stability and quality of the motion manifold. We predict a long sequence of motions by iterations in an open-loop setup, in which there is no sub-system nor feedback variables to moderate the cumulative prediction error. In each iteration, STRNN takes the last 20 frames as input and predict the next 20 frames. 

We first show results of 1000-frame prediction in the video. We also push the model further by generating longer sequences. We include motions with {\bf 20000} frames
. \figref{teaser} shows the first 2000 frames of a boxing motion. According to our experiments, STRNN can predict far longer than 20000 frames in practice, but we believe that it is already long enough to demonstrate its stability. Also, the length of prediction supersedes all other work as far as we know. The closest one is \cite{AutoCondition} but our prediction is still much longer. Also, unlike \cite{AutoCondition}, our model is trained on all the motions at once.

\subsection{Motion Denoising}
Motion capture data can be easily corrupted due to tracking errors. As a motion manifold model, STRNN can fix the data by projecting the corrupted input onto the manifold. We show the results in a controlled experiment where we have high-quality ground truth motions. It allows us to measure reconstruction errors. We employ the CMU and Edinburgh dataset. Since they contained a wide range of motions with different styles and dynamics, we randomly sampled 512 motions. Then, we randomly perturb the motion with a Gaussian noise on each dimension independently. \figref{denoiseCom} and the video show our high-quality denoising results. Numerical results such as reconstruction errors are given in \secref{comparison}, in comparison with other methods.



\subsection{Controlled Motion Synthesis}


Controlled motion synthesis is another application of STRNN. Given an initial motion and control signal, not only can we use STRNN to impose control on the given motion, the control signal can also be applied to the predicted motion too, which makes it an open-ended prediction and control. Note that although controlled motion synthesis can be done in other works (\cite{Holden:SIGGRPAH2016,Holden:SIGGRPAH2017}), our task here is much more challenging because we are applying control signals on predicted motions, which are unknown in the beginning. 

Given the initial 20 frames (or randomly chosen from the dataset), we perform prediction and control alternatively. In every iteration, there are two steps: prediction and correction. We first predict the next 20 frames. Then, we project the motion back onto the motion manifold while following the control signals explained in \secref{runtimeCosts} as much as possible. This is done by running the optimization for a number of iterations, i.e. fine-tuning the network for the current iteration. Finally, we move on to the next iteration. 

The fine-tuning is a local operation because it adapts the network only for the current iteration. Then, alternating on the open-loop prediction and the local fine-tuning can gradually drag the network out of the natural motion regions. We propose a simple solution. We back up the pre-trained weights. In run time, we use the pre-trained weights for prediction and then fine-tune the weights to satisfy the constraints. Finally, we restore the weights before the next iteration. In this way, the motion prediction is always done on the pre-trained network. In the experiments, we find that the fine-tuning only requires several iterations and thus is very quick. Since STRNN captures the motion multi-modality, the generated motion can contain multiple types of movements. Detailed results can be found in the video (long motion sequence of 1000 frames). 

\subsection{Contact Violation and Post Processing}
\label{sec:postProcessing}
In our demos, we do not use post-processing to fix foot sliding. As a result, some motions do violate contact constraints. This is partially due to the existing foot-sliding in data. We show the raw results for evaluation and comparison purposes. However, similar to existing work, post-processing such as Inverse Kinematics can be employed to fix foot sliding issues. We show one example of a controlled motion synthesis with a long and complex control trajectory. The video includes before and after post-processing results using \cite{shum_2012_realtime} 
.

\section{Comparison with Other Models}
\label{sec:comparison}
\subsection{Comparisons with Alternative RNNs}
Given there is a large body of deep learning networks in computer graphics and vision, machine learning, etc., it is impractical to do exhaustive comparisons. We, therefore, focus on the relevant area of recurrent neural networks. Some of them are designed for motion prediction such as \cite{Fragkiadaki:2015} denoted as ERD, while some of them are for action recognition such as \cite{Du:RNN:CVPR2015} denoted as BidirSimpleRNN. It is hard to do a fair black box comparison because they are designed for different purposes. Therefore, we mainly compare different spatial encoding/decoding mechanism under the same encoding/decoding/predicting setting to see how they fit data and generalize. We use the CMU dataset and keep an 80/20 training and testing ratio. We computed the same $D_{1nn}$ error explained in \secref{PredictStability}. The results are shown in \figref{modelComparison}. STRNN in general generate better motions. In addition, we found that motions generated by ERD has smaller variances than BidirSimpleRNN. We speculate that this is due to the posture encoding/decoding components in ERD. However, these components are merely fully-connected layers while STRNN models the skeleton explicitly. SRTNN still produces the best results visually as shown in the video.

\subsection{Prediction Accuracy Analysis}
\begin{table}
    \centering
    \begin{tabular}{l c c c c c c c}
        \hline
        Method  & 80 & 160 & 240 & 320 & 400 & 480 & 560\\
        \hline
        LSTM\_3LR & 0.41 & 0.67 & 1.15 & 1.50 & 1.78 & 2.02 & 2.26\\
        CRBMs & 0.68 & 1.13 & 1.55 & 2.00 & 2.45 & 2.90 & 3.34\\
        6GRAM & 1.67 & 2.36 & 2.94 & 3.43 & 3.83 & 4.19 & 4.53\\
        GPDM & 1.76 & 2.5 & 3.04 & 3.52 & 3.92 & 4.28 & 4.61\\
        ERD & 0.89 & 1.39 & 1.93 & 2.38 & 2.76 & 3.09 & 3.41\\
        \hline
        STRNN(ours) & {\bf 0.36} & {\bf 0.39} & {\bf 0.43} & {\bf 0.46} & {\bf 0.48} & {\bf 0.5} & {\bf 0.51}\\
        \hline
    \end{tabular}
    \caption{{\bf Predication Error Comparison} during the 80, 160, 240, 320, 400, 480 and 560 ms of prediction. Quantitative evaluation for longer temporal horizons is not possible due to stochasticity of human motion \cite{Fragkiadaki:2015}.}
    \label{tab:predictError}
\end{table}

Prediction accuracy has been used to compare predictive models. Although it has been shown that prediction accuracy for short-term predication is not reliable, averaging over a number of motions shows the average predication accuracy. This accuracy might not be a good metric for motion prediction of a specific type but a good indicator to show how quickly errors accumulate. We follow the protocol in \cite{Fragkiadaki:2015} and compare our model with others on H3.6M dataset \cite{h36m_pami}. We randomly choose 8 prefixes from the dataset and compare our prediction errors with some state-of-the-art methods. The error is per-frame Euclidean distance between the generated motion and the ground truth. They are: (a) 3-layer linear LSTM model (LSTM\_3LR) \cite{Fragkiadaki:2015}, (b) Conditional Restricted Boltzmann Machines (CRBMs) \cite{Taylor:2006}, (c) a nearest neighbor N-gram model with n=6 (6GRAM) \cite{Fragkiadaki:2015},  (d) Gaussian Process Dynamic Model (GPDM) \cite{Wang:PAMI2008} and (e) ERD \cite{Fragkiadaki:2015}. Also, we only compare the prediction errors within the first 560ms. This is largely due to the randomness in human motions where long predictions are not suitable for such comparisons \cite{Fragkiadaki:2015}.



We also compare our method with the one in \cite{MartinezBR_corr17} on prediction accuracy using the same per-frame Euclidean metric mentioned above. We downloaded their code and followed their protocol. Based on their code, the training and testing data come from different subjects across the same set of actions. The training data contains Subject 1, 6, 7, 8, 9 and 11. The testing data contains only Subject 5. We trained the model with the best possible setup and ran it for 200k iterations. Finally, we compute errors of the same 4 actions: Walking, Easting, Smoking and Discussion using randomly sampled motion prefixes within each action class\cite{MartinezBR_corr17}. Results are shown in \tabref{predictError2}. Numerically, STRNN performs better in all four classes. One possible reason is that the method in \cite{MartinezBR_corr17} used joint angles as representations. A small error in the joint angle space can cause big errors in the joint position space. This is consistent with our early findings when we experimented with different representations of postures. It is difficult to compare for longer duration due to the stochasticity of human motions \cite{Fragkiadaki:2015,MartinezBR_corr17}. Also, it is difficult to do visual comparison because the duration is too short. 

\begin{table*}
    \centering
    \begin{tabular}{*{1}{c}|*{4}{c}|*{4}{c}|*{4}{c}|*{4}{c}}
        \toprule
        &\multicolumn{4}{c}{Walking}&\multicolumn{4}{c}{Eating}&\multicolumn{4}{c}{Smoking}&\multicolumn{4}{c}{Discussion} \\
        milliseconds & 80 & 160 & 320 & 400 & 80 & 160 & 320 & 400 & 80 & 160 & 320 & 400 & 80 & 160 & 320 & 400 \\
        \hline
        \cite{MartinezBR_corr17} & 5.30 & 6.15 & 6.74 & 7.29 & 4.96 & 5.99 & 7.16 & 8.29 & 8.08 & 9.36 & 10.53 & 11.63 & 8.98 & 10.09 & 11.12 & 12.02 \\
        STRNN(ours) & 2.85 & 2.83 & 2.77 & 2.82 & 2.66 & 2.60 & 2.49 & 2.53 & 3.61 & 3.65 & 3.67 & 3.69 & 2.59 & 2.72 & 2.86 & 2.64 \\
        \hline
    \end{tabular}
    \caption{Predication Error Comparison during the 80, 160, 320 and 400ms of prediction for 4 actions: Walking, Eating, Smoking and Discussion.}
    \label{tab:predictError2}
\end{table*}


\subsection{Denoising Comparisons}
\label{sec:denoisingComarison}

We also compare the motion manifold qualities through motion denoising, which can be done by projecting the corrupted motions back to the manifold \cite{Holden:2015}. As STRNN does it by feeding corrupted motions into the encoders and reconstructing denoised motion in the decoders, the reconstruction error is a good way to evaluate the manifold quality. We compare our method with \cite{Holden:2015} (SC), using the source code downloaded from the authors' website. The CMU and Edinburgh (CE) dataset are used because of their high quality.
We randomly selected 512 motions. To show how our method is far more robust against stepping pattern noises, we use Gaussian noises with zero mean and 0.3 standard deviation for all joint positions, and 3 and 5 standard deviations on the stepping patterns to generate two polluted datasets. Note that the standard deviation of the noise added onto the feet joints is still 0.3 (i.e. the same as any other joints). It is only the stepping patterns, which are four binary variables, are disturbed with noises with higher standard deviations (3 and 5). In both implementations, for the stepping pattern parameter, a value higher than 0.5 is regarded as one or zero otherwise.

\begin{figure}
    \centering
    \includegraphics[width=1.\linewidth]{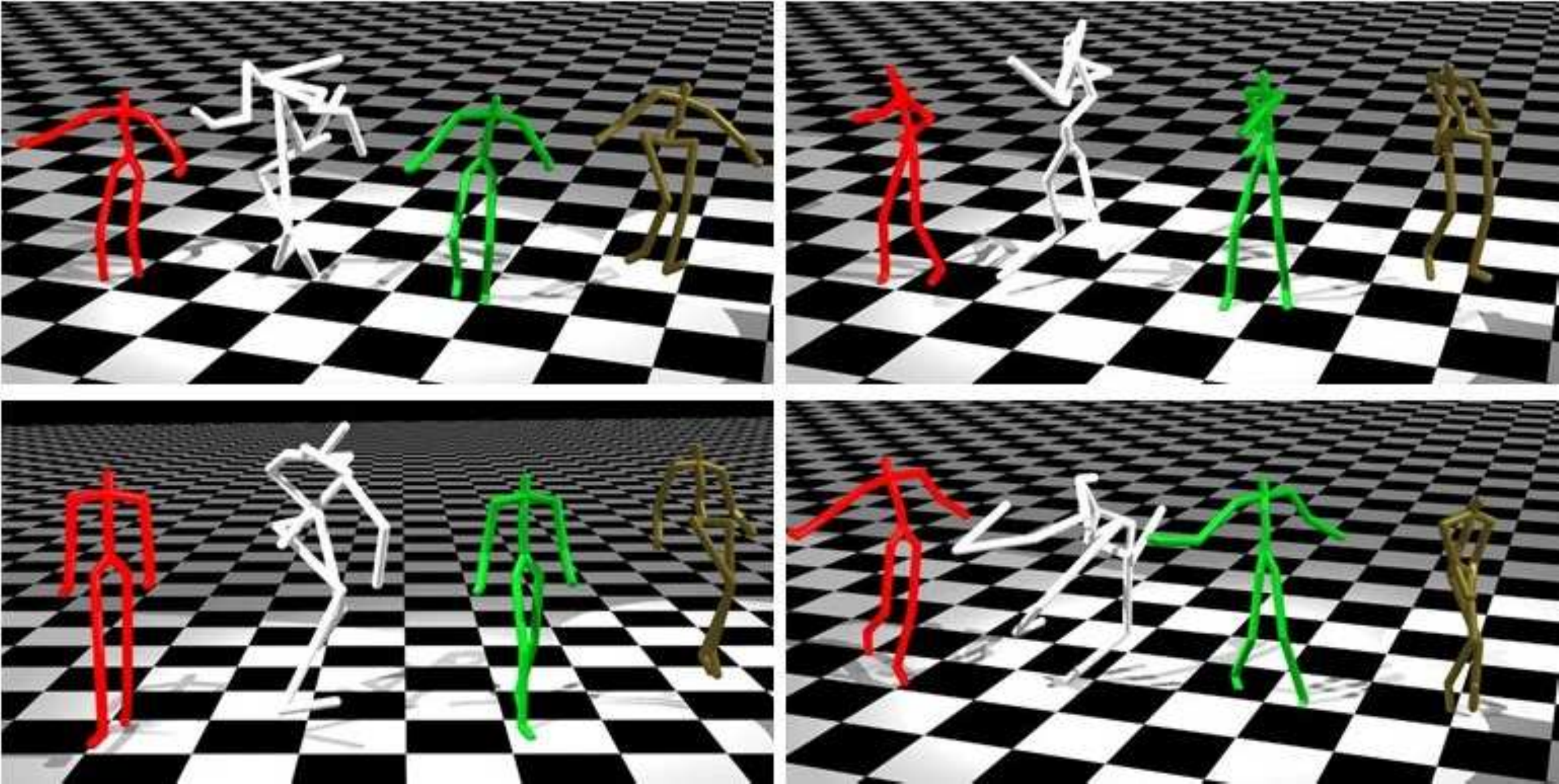}
    \caption{Denoising comparisons. Red: original, White: corrupted, Green: STRNN, Yellow: SC}
    \label{fig:denoiseCom}
\end{figure}

Four sets of screenshots of the corresponding frames in the original, corrupted and denoised motions are shown in \figref{denoiseCom} and the video. The method proposed in \cite{Holden:2015} tends to overly smooth motions. We speculate that this is mainly due to the under-estimate of motion variances as their do convolutions and max-pooling along time. In contrast, STRNN captures the dynamics of the motions well. Also, SC depends heavily on precise clear stepping patterns. For motions that involve unclear stepping patterns, such as dancing with intentional foot skate, or motions with corrupted stepping patterns, their method may not be able to produce reasonable results.

\begin{figure}
    \centering
    \includegraphics[width=0.98\linewidth]{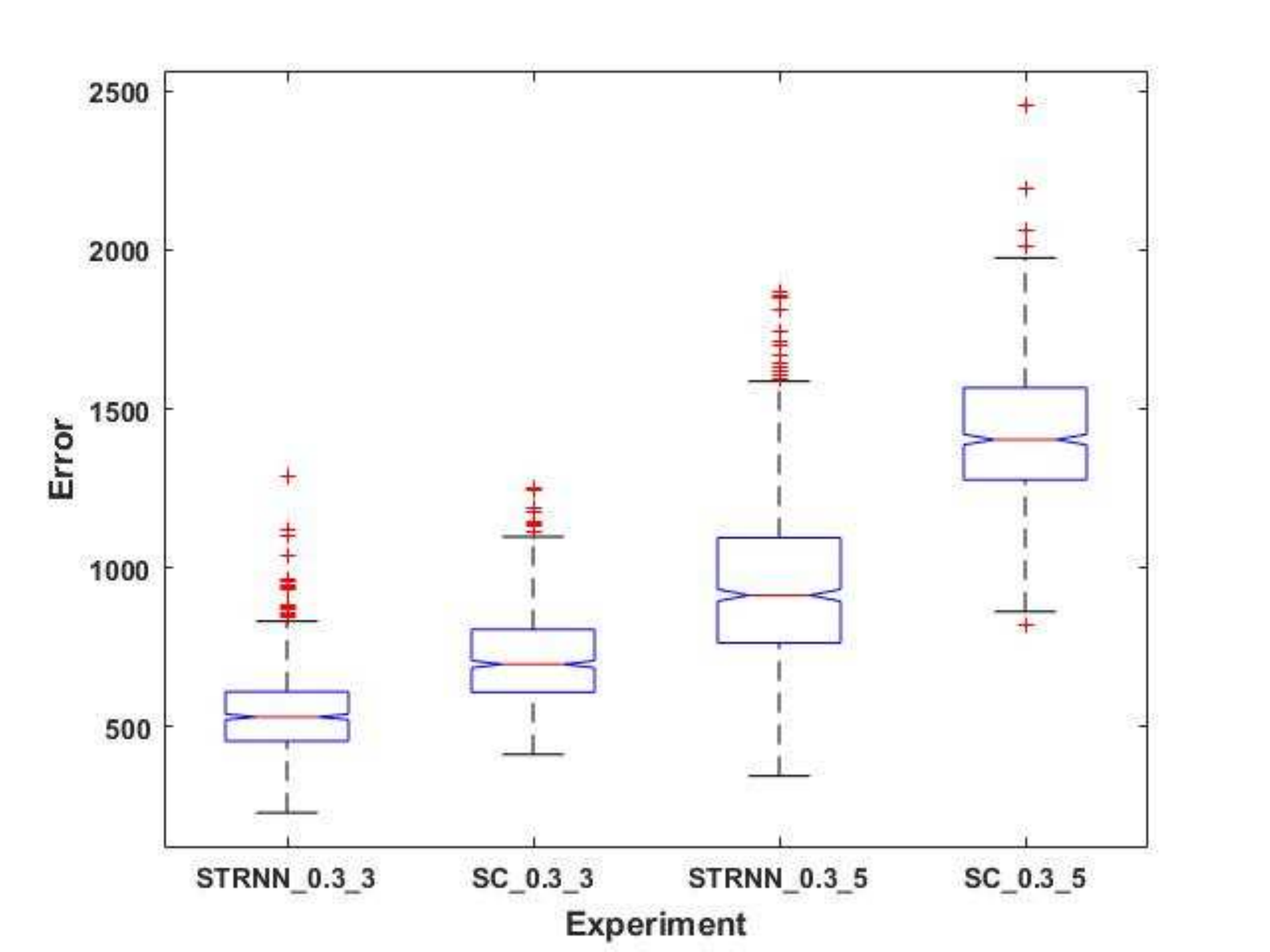}
    \caption{X Axis: Experiments. Y Axis: Reconstruction error of STRNN\_a\_b and SC\_a\_b \cite{Holden:2015}. a and b are the standard deviations of the Gaussian noise patterns added on the body motion and stepping patterns.
    }
    \label{fig:denoiseErrorComparison}
\end{figure}

For numerical comparison, we consider the sum of squared errors of the joint positions of all motions between the original and denoised motions, as summarized in \figref{denoiseErrorComparison}. 
The reconstructed motions by both STRNN and SC are denoted as STRNN\_0.3\_3, SC\_0.3\_3, STRNN\_0.3\_5 and SC\_0.3\_5 respectively. 
Given the same noise level, STRNN's reconstruction error is smaller than SC's. Also, when the noise level becomes bigger on the stepping patterns, STRNN is less influenced than SC is.

\section{Performances}
\label{sec:performance}
The code, implemented with Theano and Keras, runs on a computer with Intel(R) Xeon(R) CPU E5-1650 v4 3.6GHz, 64GB memory and an Nvidia GTX 1080 graphics card. For motion manifold learning, the pre-training on CMU and Edinburgh dataset takes approximately 16 hours. Once the network is trained, the motion generation is fast. For long-horizon extrapolation, the system generates motions at 80000 FPS on average. For motion synthesis, we run 20 steps to satisfy various constraints for every 40-frame window. The system still generates motions at 1200 FPS on average. 
\section{Discussion}
\label{sec:discussion}

\subsection{Alternative Architecture of STRNN}

Firstly, both the TDecoder and TPredictor can be \textit{conditional} or \textit{unconditional}. During training, an unconditional TDecoder takes the ground truth $X_t$ for decoding $X_{t-1}$ while a conditional decoder takes the reconstructed $\hat{X}_t$ for decoding $X_{t-1}$. The same logic applies to TPredictor. 
The conditional scheme allows the model to learn multiple modes in the data distribution, without which, the model might average multiple modes. However, if data correlation is big (i.e. consecutive frames are too similar to one another), the conditional scheme forces the model to learn the correlation so that the temporal motion variance is under-estimated. Furthermore, the gradient for fixing the error is small and it requires long-term knowledge about the input sequence. In our context, all these concerns are not existent. We choose the conditional scheme because human motions are multi-modal. The design of STRNN enables us to use the long-term knowledge of the input sequence so that if high correlation does appear, there will be a gradient to fix it. However, if required, it is possible for STRNN to generalize further by using either an unconditional scheme or a hybrid approach (e.g. unconditional TDecoder and conditional TPredictor or vice versa).

Next, the temporal memory mechanism can be modeled by other neural cells such as Gated Recurrent Unit (GRU) \cite{Cho:MBB14}. \cite{ChungGCB:CoRR2014} shows a detailed comparison between LSTM and GRU on many datasets, and finds no conclusive evidence to prefer one to the other. We found that LSTM performed better by different margins on the datasets used. We speculate that this is due to the memory unit in LSTM that keeps the mid/long horizon dependencies.

There are other strategies regarding decoding/predicting mechanism in RNNs for different tasks. The Residual Network structure \cite{MartinezBR_corr17} facilitates the learning because all the decoder/predictor needs to learn is the first-order motion information. The Attention mechanism \cite{IonescuSminchisescu11} makes use of the whole prefix as a context for decoding, such that the decoder/predictor would be more context-aware. We experimented with a model similar to \cite{MartinezBR_corr17} gave lower training errors but generated results similar to LSTM\_3LR (i.e. converging to a mean pose in long-horizon prediction).
We also tried introducing the Attention mechanism, which gave worse training/testing errors and generated highly jittering motions in long-horizon prediction. We speculate that this is because all the frames in the motion prefix are taken as a context while some of them should not have influence on later frames at all.

The multi-objective control (MOC) method in \cite{Lee:SIGGRPAHAsia2018:CharCtrl} can also generate high-quality motions under user controls. However, our method is orthogonal to MOC in several aspects. It is therefore not very meaningful to do a direct numerical comparison. STRNN aims to learn a global natural motion manifold by learning all unlabeled motions together while MOC learns to control specific labelled actions. Also, MOC learns controlled motion transitions through manually crafted motion grammars while STRNN learns transitions purely from data and randomly generates them in an open-loop setting.

STRNN uses a spatial encoder to project the frames onto a higher-dimensional space (512 hidden units), similar to \cite{Fragkiadaki:2015,MartinezBR_corr17}, which is somewhat counter-intuitive and different from previous findings that human motions can be embedded into a lower-dimensional space using techniques such as Principle Component Analysis \cite{Safonova:2004:SPR:1015706.1015754}. We tried several different settings to map motions onto a lower-dimensional space, such as reducing the cell number is L1-L4 in \figref{spatialNet}. We found that they are all unable to learn the motion manifold well. While this does not mean deep neural networks in lower-dimensional space is unsuitable for the problem, our empirical evidence shows that it could be challenging for them to capture the temporal multi-modality.

Lastly, the temporal network can consist of more than one layer. Stacking multiple layers of LSTM can model deeper temporal non-linearity in the data. In our dataset, it is not required since joint trajectories can be well approximated as piece-wise linear functions. However, it could be needed for other data types such as texts and videos.

\subsection{Limitations}
There are some limitations with STRNN. The transitions in the training data are important. During extrapolation, it is easier for STRNN to transit from one motion to another when there is at least one motion coming next. Essentially it has to do with the connectivity of the data. Given the sheer volume of the training data and that nearly all captured motions end somewhere close to a neutral standing posture, it is almost always the case. More training data can also improve the situation. 

Second, when multiple next motions are possible, the transition can take a number of frames to transit into the next motion, which creates a bit of ``hesitation'' in the motion. It is caused by the roughly equal probabilities of several candidate motions. The equilibrium is very soon broken by the accumulated error that acts as a perturbance that brings the system into favoring one candidate motion. This can also be caused by the rest poses in the training data where the subject does not move much for a period of time, as noticed as ``dead-times'' in \cite{AutoCondition}. 

In addition, since our model does not take the environment information into consideration, with training data such as climbing a ladder or stepping on/over some obstacles, our model sometimes generates those motions too. In the future, we wish to incorporate environmental information into the network.

Lastly, since there is no motion labelling such as ``jumping'' or ``walking'', there is no global control in the action level. A direction to be pursued is to incorporate action labels. One possible approach is to convert STRNN into a conditional network where the motion generation is conditioned on action labels. A related direction is a global control mechanism that enables the users to specify actions at particular timings.

\section{Conclusion}
\label{sec:conclusion}
In this paper, we propose a new deep-learning framework to learn a motion manifold. Comparing to existing systems, our framework creates better results in various applications. The success is built upon two innovative solutions to maintain motion variances. We construct a hierarchical spatial encoder by dividing the skeletal structure into parts with strong spatial proximity, thereby preserving motion variances. We also construct a batch prediction network that allows long-horizon optimization, thereby harnessing the motion variances.

\section*{Acknowledgments}
The project is partially supported by EPSRC (Ref:EP/R031193/1). The authors wish to gratefully acknowledge the support of NVIDIA Corporation with the donation of the Titan Xp GPU used for this research. The project was supported in part by the Royal Society (Ref: IES$\backslash$R2$\backslash$181024).

\bibliographystyle{IEEEtran}
\bibliography{references}

\vspace{-3em}
\begin{IEEEbiography}[{\includegraphics[width=1in,height=1.2in,clip,keepaspectratio]{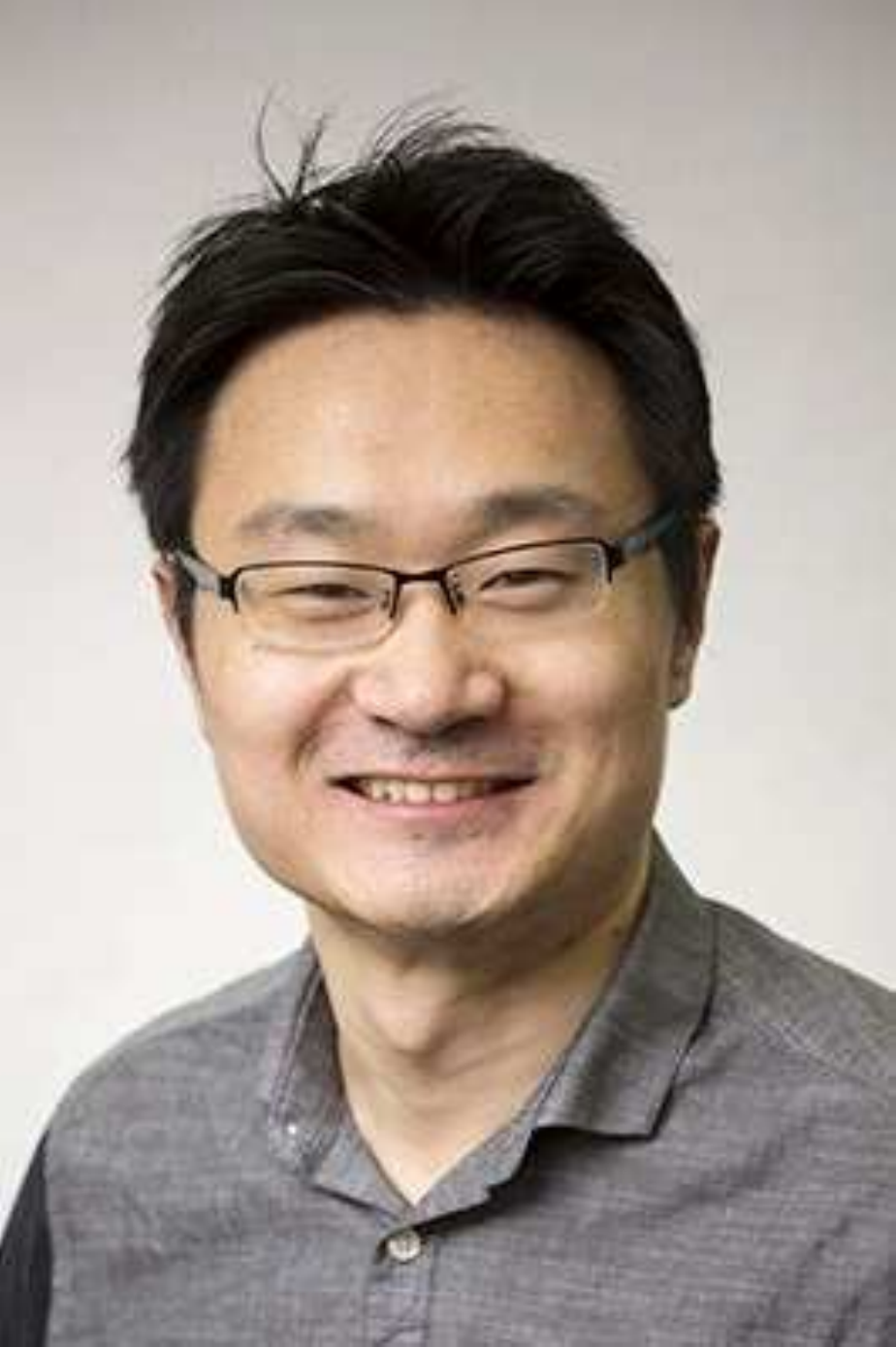}}]{He Wang}
is an Assistant Professor (Lecturer in UK) at School of Computing, University of Leeds, UK. His research interest is computer graphics, vision and machine learning and applications. Previously he was a Postdoctoral Associate at Disney Research Los Angeles. He received my PhD in 2012 and did a post-doc afterwards both in School of Informatics, University of Edinburgh. Before his PhD, he worked in industry for 4 years after graduating from Zhejiang University, China.
\end{IEEEbiography}

\vspace{-4em}
\begin{IEEEbiography}[{\includegraphics[width=1in,height=1.1in,clip,keepaspectratio]{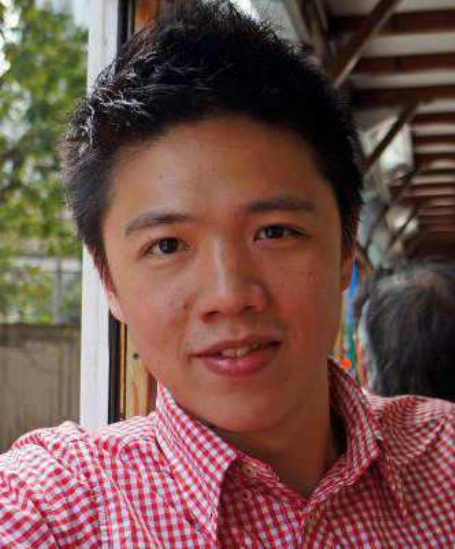}}]{Edmond S. L. Ho}
is a Senior Lecturer in the Department of Computer and Information Sciences at Northumbria University, Newcastle, UK. Before joining Northumbria University in 2016, he was a Research Assistant Professor in the Department of Computer Science at Hong Kong Baptist University. He received the BSc degree from the Hong Kong Baptist University, the MPhil degree from the City University of Hong Kong, and the PhD degree from Edinburgh University.
\end{IEEEbiography}
\vspace{-4em}
\begin{IEEEbiography}[{\includegraphics[width=1in,height=1.1in,clip,keepaspectratio]{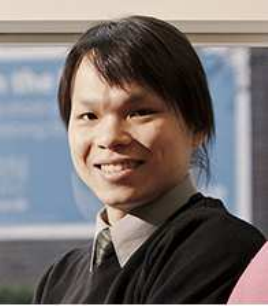}}]{Hubert P. H. Shum}
is an Associate Professor (Reader) in Computer Science at Northumbria University and the Director of Research and Innovation of the Computer and Information Sciences Department. Before that, he was a Senior Lecturer at Northumbria University, a Lecturer at the University of Worcester and a postdoctoral researcher at RIKEN Japan. He received his PhD degree from the University of Edinburgh, his Master and Bachelor degrees from the City University of Hong Kong. 
\end{IEEEbiography}

\vspace{-3em}
\begin{IEEEbiography}[{\includegraphics[width=1in,height=1.1in,clip,keepaspectratio]{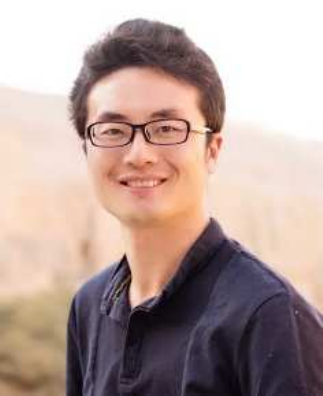}}]{Zhanxing Zhu}
is an Assistant Professor at Peking University and Beijing Institute of Big Data Research. He is closely affiliated with Deep Learning Lab of Peking University. Previously he obtained his PhD in machine learning from School of Informatics of University of Edinburgh, UK. His research interests cover methodology/theory of machine learning and artificial intelligence and their applications in various areas.
\end{IEEEbiography}
\vfill
\end{document}